*Title:*

Spatial Structure and Scaling of Agricultural Networks

*Author names and affiliations:*

D. Sousa[1*] and C. Small[1]

[1]Lamont-Doherty Earth Observatory, Columbia University, Palisades, NY 10964 USA

Email: d.sousa@columbia.edu. Tel.: +1 5303044992

*Abstract:*

Considering agricultural landscapes as networks can provide information about spatial connectivity relevant for a wide range of applications including pollination, pest management, and ecology. Global agricultural networks are well-described by power law rank-size distributions. However, regional analyses capture only a subset of the total global network. Most analyses are regional. In this paper, we seek to address the following questions: Does the globally observed scale-free property of agricultural networks hold over smaller spatial domains? Can similar properties be observed at kilometer to meter scales? We analyze 9 intensively cultivated Landsat scenes on 5 continents with a wide range of vegetation distributions. We find that networks of vegetation fraction within the domain of each of these Landsat scenes exhibit substantial variability – but still possess similar scaling properties to the global distribution of agriculture. We also find similar results using a 39 km$^2$ IKONOS image. To illustrate an application of spatial network analysis, we show an example of network disruption. We compare two networks with similar rank-size distributions that behave differently when nodes are progressively removed. We suggest that treating agricultural land cover as spatial networks can provide a straightforward way of characterizing the connectivity of complex spatial distributions of agriculture across a wide range of landscapes and at spatial scales relevant for practical agricultural applications.

*Introduction*

The spatial distribution of agriculture in a landscape can provide information which is complementary to the properties of individual fields or political units. Pollination, insect diversity, and other ecosystem services are reliant on the spatial connectivity of an agricultural landscape (Diekötter et al. 2008; Ricketts et al. 2006). Outbreaks of pests and pathogens can sometimes be contained by breaking spatial adjacency between cropped areas (Gilligan 2008). The ecology of native species populations can be altered by habitat fragmentation of natural landscapes by agriculture (Dixo et al. 2009; Luoto et al. 2003). However, the diversity of agricultural landscapes around the globe demands a metric which is flexible enough to accommodate a wide range of spatial distributions and connectivity patterns. Networks are one conceptually simple tool which can represent a variety of processes with complex spatial structure.

Globally, estimates of cropland have been observed to display an unexpected consistency in their size distributions (Small and Sousa 2015a). Despite considerable disagreement when compared directly in the same locations, 4 different global agriculture products possess the property that the sizes of contiguous patches of agricultural land diminish at the same rate that their frequency increases (Figure 1). This property implies (nearly) uniform distributions of total agricultural area across scales – the sum of the area of the largest segments is equal to the sum of the area of the smallest segments, which is equal to the sum of the segments of any arbitrary size interval in between. The consistency of this observation across the 4 products is especially surprising given the substantial differences in the input data, assumptions, and algorithms used in each of the 4 products. The consistency of the observation at global scales begs the question of whether this pattern can also be observed at finer spatial scales.

The property of diminishing magnitude with increasing number is common in nature and is often referred to as a power law relationship. Power law relationships are also a defining characteristic of many networks – often referred to as "scale-free" because of the implied self-similarity. Because networks are capable of representing processes with complex spatial structure, and because many networks display similar power law relationships to those observed for agriculture on the global scale, we suggest that networks may be a useful tool to characterize agricultural landscapes and provide insight into processes reliant on agricultural connectivity.

The objective of this paper is to investigate the question of whether this scaling property holds over smaller areas and at spatial scales relevant to the questions of pollination, pathogen transmission, and ecology. Specifically, we seek to assess the robustness of the global scaling relationship and ask the question: Do the size distributions of agricultural landscapes maintain similarity to true power laws over smaller spatial domains?

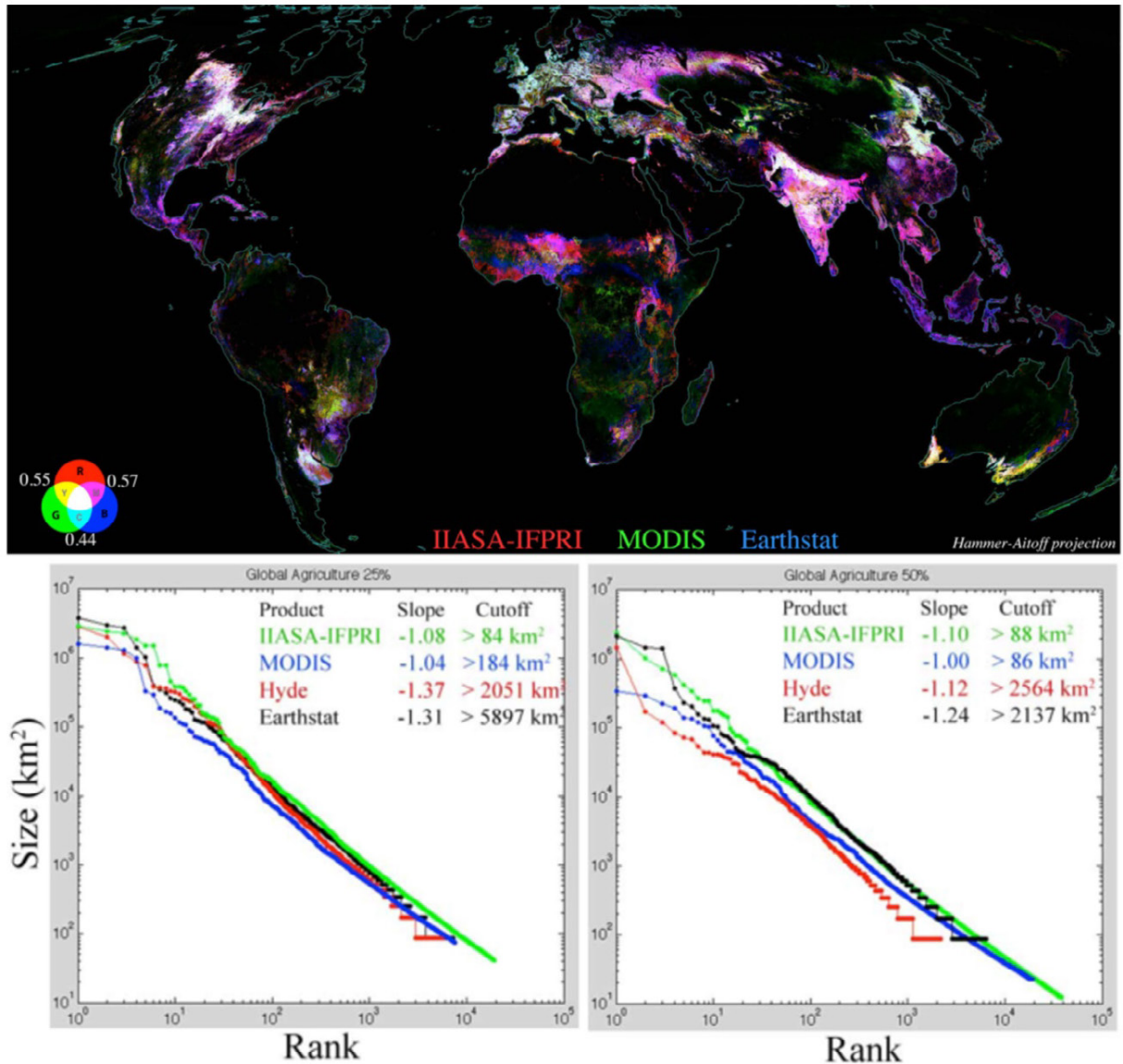

Figure 1. Global comparison of agricultural land cover products (top) and corresponding rank size distributions of agricultural land area (bottom). Areal fractions of land under cultivation for three global products shown as red, green, and blue brightness. Pairwise spatial correlations between products shown in lower left corner of the map quantify agreement. Segmenting each continuous fraction map at >25% and >50% thresholds produces binary maps. Rank size distributions of contiguous areas of agricultural land cover for each product have similar slopes over 4 orders of magnitude in size. Power law fits to each rank size distribution yield slopes near -1. Size cutoffs estimate lower bound of power law behavior. Small cutoffs for the IASA-IFPRI and MODIS products indicate that the power law fits all but the smallest segments, while the larger cutoffs in the HYDE and Earthstat products result from quantization of smaller segments due to their coarser 10 x 10 km resolution. Slopes near -1 indicate that areas of agricultural land cover diminish in size at the same rate they increase in number. The implication is that the total area of agricultural land cover is nearly uniform across a wide range of segment sizes.

*Background*

### a. Rank-size Plots and Long-Tailed Distributions

Some processes in nature tend to cluster around one common value, with large deviations from this value being relatively infrequent. However, other processes can take on a wide range of values – sometimes varying over several orders magnitude. When viewed as realizations of random variables, the probability distribution of a random variable which takes on a wide range of values is said to be heavy tailed. In a heavy tailed distribution, concepts from Gaussian statistics such as mean and standard deviation have little utility since the random variable deviates highly from that of a Gaussian (i.e., extreme events are much more common than predicted by a Gaussian distribution). Several types of heavy-tailed distribution which have been invoked by different authors to describe natural phenomena include the Weibull (e.g. wind speed, (Seguro and Lambert 2000)), log-normal (e.g. distribution of chemical concentrations, (Limpert et al. 2001)) and power law (e.g. city size, (Auerbach 1913; Lotka 1941; Zipf 1942)) distributions.

In the case of phenomena characterized by heavy tailed distributions, rank-size plots can be an intuitive tool for displaying both the magnitude and frequency of observations. Because such processes often span several orders of magnitude, such plots are typically displayed on logarithmic axes. Such a visualization scheme can be desirable because of its conceptual simplicity and minimum of assumptions about the form of the data. In the case where a rank-size plot is linear on logarithmic axes, the power law distribution is often considered a likely candidate for the underlying process. A power law distribution is defined by a constant factor and an exponent. If a set of features is distributed according to a power law, the slope of the rank-size plot in log-log space is related to the power law exponent α by the following expression (Li 2002):

$$slope = -\frac{1}{\alpha - 1}$$

Nonparametric statistical methods have been developed to determine the power law of best fit, the portion of the distribution most likely to be power law, and confidence intervals using Monte Carlo methods and the Kolmogorov-Smirnoff goodness-of-fit statistic. For an excellent description of these methods and application to a wide range of datasets, see (Clauset et al. 2009). When observations are binned logarithmically, a rank-size distribution with a slope of -1 (corresponding to a power law exponent of -2) corresponds to a uniform distribution across scales (Small et al. 2011).

Importantly, linearity of the rank-size plot alone does not rule out the possibility of other similar heavy tailed distributions describing the data equally well – or even better (Clauset et al. 2009). For this reason, in this paper we only use power law fitting as a convenient way to quantify the degree of linearity and slope of the rank-size plots. We remain noncommittal about the ultimate form of the underlying probability density function and suggest more rigorous analysis as a direction for future work on this topic.

### b. *Scale-Free Networks and Constrained Networks*

The most basic pieces of networks are *nodes* and *links*. Nodes are connected to each other by links. Depending on the network, some nodes may be linked to many other nodes, some may be linked to only a few, and some nodes may not linked to any other nodes at all. Each set of interconnected nodes is called a *component*. Within each component, all nodes are connected to each other either directly or indirectly (i.e. through other nodes within the same component). No node within one component can be linked to a node within another component. A *network* is a set of components. In many networks, all nodes are linked to each other (directly or indirectly) to form a single component (Newman 2010). Other networks have many components.

In a network, each node has a certain number of links. The distribution of the number of links per node is called the degree distribution of the network. In some networks, the degree distribution can be characterized by a power law. These networks are called scale-free networks. For these networks, when the distribution of degree sizes versus rank (ordinal number) is plotted on logarithmic axes, the result is linear. The slope of this line can vary substantially for different networks (Barabási and Albert 1999). The wide range of degrees necessary for a power law distribution is possible in some cases because many networks have no limit (or some very large limit) to the number of links that each node can have. Networks are already used in the field of landscape ecology (Cantwell and Forman 1993; Gardner et al. 1992; McIntyre et al. 2014; Urban and Keitt 2001) under the term graph theory. For a general review of network theory, see (Albert and Barabási 2002) and (Newman 2010).

In this paper we treat landscapes as networks of land cover. The spatial domain of interest determines the total possible spatial extent of the network it contains. In this paper, each pixel is treated as a potential node. A pixel becomes a node of a spatial network if it satisfies the criteria for existence. In this paper, we have a single criterion: subpixel vegetation abundance above the threshold of analysis. We consider two pixels to be directly linked if they are spatially adjacent to each other. For this reason, nodes in land cover networks as considered here have a maximum number of direct links (Steinwendner 2002). Because we use the Queen's case for connectivity (all immediate neighbors including diagonals), this number is 8. In this case, the parameter of interest is not the degree distribution but the component size distribution, as the sizes of each component (spatially contiguous patch of agricultural land) can possess a wide range of values. The rank-size plots shown in this paper show the distribution of components in a single network. We refer to the particular type of spatial network defined in this way as a *bounded spatial network* (Small and Sousa 2015b).

(Small and Sousa 2015a) show that four land cover products which seek to map agriculture at the global scale exhibit empirical component size distributions characterized by linearity in logarithmic space and slope of -1, despite differences in spatial patterns (Figure 1). This result holds across a wide range of analysis thresholds (described in more detail below). This suggests that agriculture may be well characterized as a scale-free spatial network on the global scale. Other spatially continuous fields have also been found to exhibit similar properties on the global scale and are discussed in detail in (Small and Sousa 2015b).

Scale-free networks have been shown to result from simple conditions: network growth and preferential attachment (Barabási and Albert 1999). Preferential attachment is the tendency for new nodes to attach more frequently to existing nodes with greater numbers of links (or to components with a greater number of nodes) than to their less connected counterparts – the rich get richer. The networks we consider fill space on a surface. This generates a mechanism for preferential attachment because the surface has finite area and larger components have larger perimeters to which new nodes can link. If new nodes are generated randomly in space, components with larger perimeters will exhibit preferential attachment – without the need for a specific mechanism for preference. To the extent that components with larger sizes (i.e. areas) also have larger perimeters, a mechanism for preferential attachment is then inherent to bounded spatial networks on a surface. For more detailed background and mechanism, see (Small and Sousa 2015b).

*Data & Methods*

To quantify the scaling properties of different agricultural landscapes, we choose images that are dominated by agricultural land cover and then use the following procedure. Beginning with raw Landsat data, we first calibrate from DN to radiance to exoatmospheric reflectance. We then estimate vegetation fraction ($F_v$) at each pixel using the standardized global endmembers from (Small and Milesi 2013), generating a continuous field of sub-pixel vegetation abundance. We then segment the $F_v$ images at several different fraction thresholds with the ENVI segmentation algorithm, using the Queen's case of 8 neighbors including diagonals. We use the Queen's case in order to provide the most liberal estimate of connectivity. We use a minimum segment size of 9 to account for spatial autocorrelation of the input imagery and avoid large numbers of spurious detections. The segmentation algorithm produces a map of segments corresponding to spatially contiguous patches of vegetation (for each threshold). Next, we calculate the total area of each segment (for each threshold). The resulting segment size maps (for each threshold) provide both the size distributions and a depiction of the spatial network structure. Segment areas are then sorted into a descending list and plotted against ordinal number (i.e. rank) on logarithmic axes.

A wide range of thresholds was applied in each case and results were compared. Figure 2 shows the typical progression of a rank-size distribution at full resolution for an example region in northern California. Images of the spatial structure of the network are shown for several different thresholds, with inset size distributions. Segment sizes are color coded on both the image and the rank-size plot. At threshold equal to 100% subpixel vegetation abundance, all pixels fall below the threshold and there is no network. As the threshold is lowered, more pixels are included in the network and form components (contiguous patches). At this phase the components correspond to individual fields or groups of closely spaced fields with high $F_v$. Continuing to lower the threshold eventually results in connection of more and more components into larger contiguous areas as pixels with lower $F_v$ are added to the network. Eventually enough pixels become part of the network that they superconnect and form one massive unit. If the threshold continues to be lowered to negligible $F_v$, the entire spatial domain of the image becomes part of the network. For more detail on the general methodology used to segment continuous fields, see (Small and Sousa 2015b) and (Small et al. 2011).

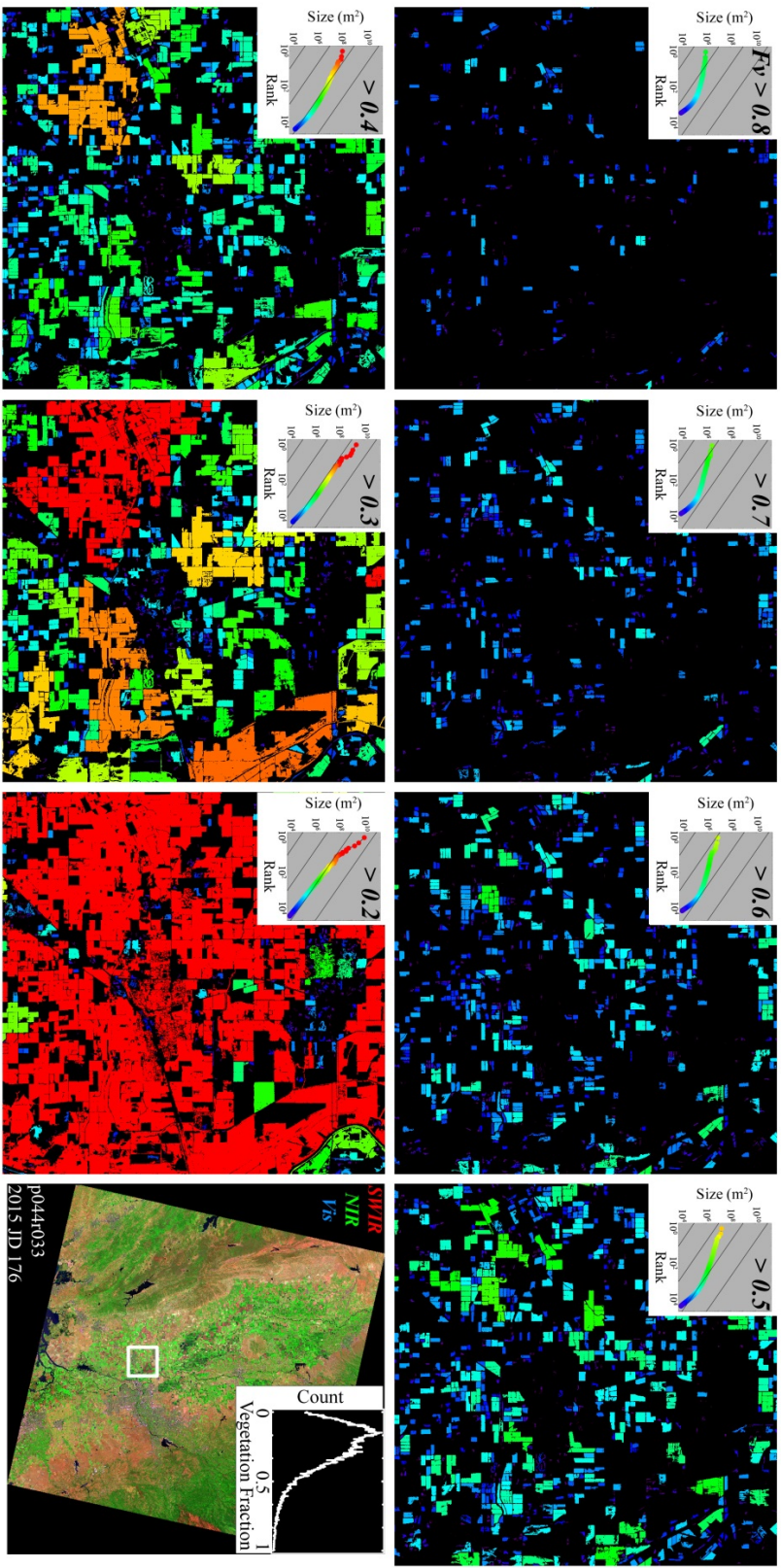

Figure 2. Illustration of network progression with threshold. This 1000 x 1000 pixel subscene of Landsat vegetation fraction was thresholded at successive values (upper left of each image). Insets show size distributions for each threshold. At threshold of Fv > 0.8, only a few fields emerge as part of the network and the size distribution is nearly flat. As the threshold decreases, more fields are included and adjacent fields connect. The size distribution steepens and loses its curvature. This continues until the size distribution becomes straight (near Fv > 0.3 in this case). After this point, further lowering of the threshold results in a majority of segments superconnecting into a small number of very large segments. This progression is typical of the other vegetation fraction images in this paper.

Disruption of agricultural networks was performed by sequential erosion using a morphological operator. For each iteration of the analysis, segment area maps were converted into binary maps indicating presence or absence of agriculture. These binary maps were then convolved with a 3 x 3 pixel Gaussian filter. Any pixel with a full set of 8 agricultural neighbors was unchanged, but any pixel with one or more non-agricultural neighbor decreased in value. A threshold of 1 was then applied and segment areas were recaluclated. This produced the effect of removing every pixel in the image on the boundary of the network. The output of one erosional step was then used as the input for the next step.

Power law exponents were fit using the statistically robust algorithm described in (Clauset et al. 2009) and converted to slopes of the size distributions using the relation given above. Power law fits are also characterized by size cutoffs describing how far the power law properties plausibly extend down the lower tail of the distribution. Cutoffs were determined using the same algorithm by choosing the minimum of the Kolmogorov-Smirnoff (KS) statistic for sets of points extending sequentially farther into the lower tail of the distribution. Significance was estimated using a Monte Carlo approach to generate 1000 synthetic datasets and calculating the KS goodness-of-fit for each. Using this approach, large p values represent plausible power law fits. We use the suggestion of (Clauset et al. 2009) in presenting significant power law fits as those with $p > 0.1$, which is a stricter test than accepting as plausible distributions with $p > 0.05$. While significant p values indicate that a power law distribution cannot be ruled out, they do not decisively show it to be a better fit than other heavy tailed distributions such as log-normal.

The data used in this study were (1) Landsat TM/ETM+/OLI scenes selected from 9 diverse agricultural regions across 5 continents and (2) one 7.4 x 5.2 km IKONOS scene of an intensively cultivated region in Anhui, China. All Landsat scenes were acquired from the USGS Earth Resources Observation and Science Center (www.glovis.usgs.gov). The scenes were chosen to represent a diverse set of landscapes dominated by extensive agriculture, spanning a range of field sizes, climate zones, phenologies, and land management practices. A wide range of crops are represented, including regions dominated by one or two grains (e.g. rice and/or wheat) as well as regions producing a balance of both commodity and specialty crops. In all cases, we use UTM equal area projections at the native resolution of the sensor. Landsat scenes in this analysis are referred to by their WRS-2 path and row identifiers: i.e. scene p029r030 corresponds to Path 29, Row 30 (South Dakota).

Fig 3a shows false color composites of the 9 Landsat scenes used for this analysis. Spatial configuration of agriculture varies widely from nearly wall-to-wall coverage (e.g. p029r030, p123r035) to regions strongly limited in spatial extent by irrigation (e.g. p039r037, p151r038). A range in extent of sectioning of the landscape by roads and rivers is apparent. Field size varies widely both across scenes and within scenes. All scenes contain some non-agricultural vegetation ranging from tropical forest to desert shrubs – but all are dominated by agriculture. The Bavaria scene contains several forest patches, but all are managed forests so are effectively part of the agriculture/silviculture mosaic. Scenes were chosen at varying stages of the annual cycle, from soon after planting to maximum greenness. The spatial extent and abundance of this vegetation varies from scene to scene. While the presence of some non-cultivated vegetation violates the assumption made in the analysis that networks of vegetation fraction strictly represent networks of agricultural activity, we have attempted to choose regions dominated by extensive cropland. We also suggest that, for some applications such as species migration and pollination, vegetation networks may be closer to the phenomenon of interest than strict definitions of cropland. Further, while considerable uncertainty exists as to the definition of cropland in global agriculture maps (reviewed in (Small and Sousa 2015a)), subpixel vegetation abundance represents a physically meaningful quantity which can be directly compared across widely varying landscapes. While using $F_v$ as a general proxy for agriculture would not be valid in many landscapes, we hold that its properties of simplicity and consistency justify its use in the examples chosen in the context of this analysis.

## *Analysis*

### *a. Landsat*

Fig 3b shows Rank-Size plots for 3 different thresholds for each of the 9 Landsat scenes from Fig 3a. $F_v$ distribution for each scene is inset with the three thresholds indicated using vertical arrows. Histograms vary widely from scene to scene in central tendency, dispersion, and number of modes, reflecting differences between the landscapes described above. Thresholds are adjusted accordingly from scene to scene to capture similar positions in the distribution. Horizontal arrows on the rank-size plots indicate the cutoff for power law fit that maximized the goodness-of-fit criterion. Italicized thresholds and slopes have p values > 0.1, indicating a statistically plausible power law fit. The statistical significance of the fit is not critical for the purposes of this analysis because we use the power law exponent as a tool to quantify the slope of the rank-size plot - not as an assertion of the generating process itself. We include the goodness-of-fit result for the benefit of readers inclined to favor the power law mechanism.

Fig 3c shows Rank-Size slope estimates for several thresholds for each of the 9 Landsat scenes. Error bars indicate 95% confidence. As the threshold is successively lowered, rank-size slopes generally increase toward more negative values. This corresponds to an increase in overall network size and in the size of individual components, consistent with the network growth mechanism proposed in (Small and Sousa 2015b). Prominent exceptions to this rule correspond to cases of severe non-Gaussianity of the vegetation histogram, e.g. bimodality in Landsat scenes p224r067 and p123r036 and a broad, asymmetric shoulder in scene p029r030. Slopes near -1 indicate that segments decrease in size at roughly the same rate that they increase in frequency. Slopes pass through a value of -1 for 8 of the 9 scenes considered here. The two scenes with slopes consistently shallower than -1, p151r038 and p039r037 are characterized by exponential-like $F_v$ histograms with a mode of $F_v \approx 0$.

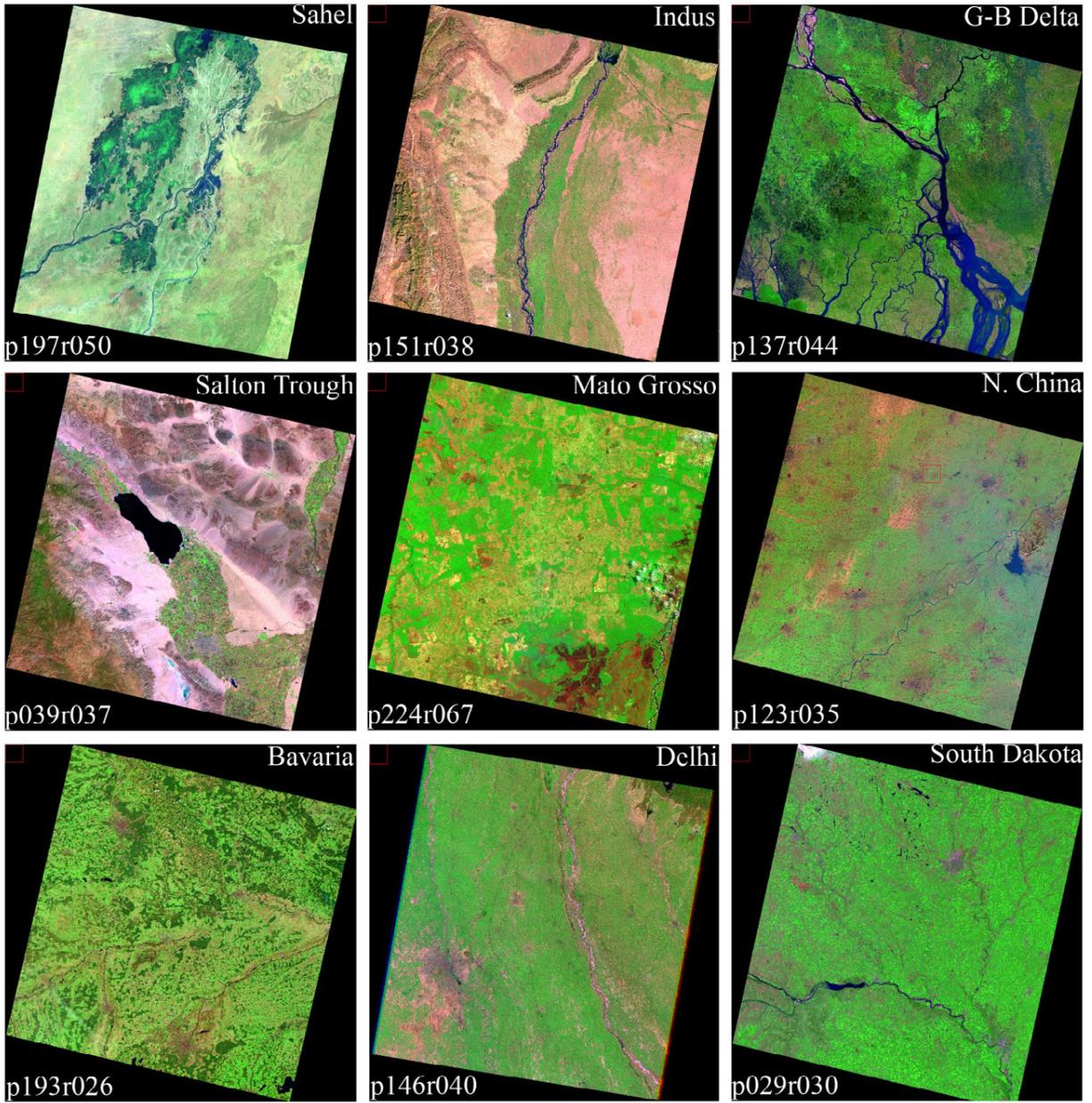

Figure 3a. Agricultural landscapes used for scaling analysis. Scenes were chosen to represent a diverse set of landscapes characterized by agricultural extensification and intensification. A range of field sizes, competing land uses, climate zones, and land management practices is depicted.

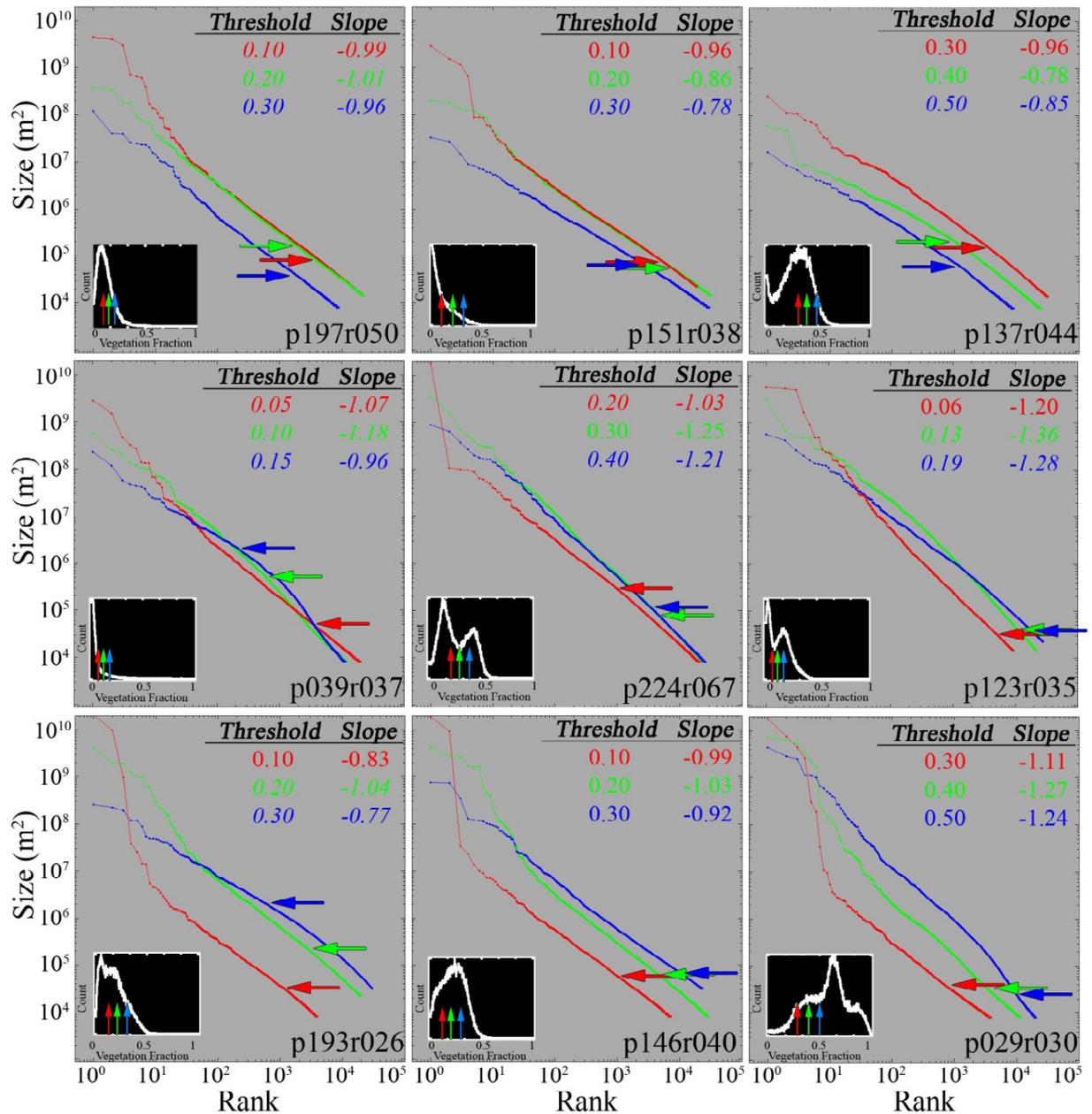

Figure 3b. Rank-Size distributions for vegetation fraction from the 9 Landsat scenes shown in Figure 3a. Inset shows vegetation fraction histogram for each Landsat scene, with arrows indicating the segmentation thresholds. Rank-Size distributions for each scene illustrate the sensitivity of the network structure to threshold. Distributions of vegetation fraction are different for each landscape but most scenes have linear rank size distributions with slopes near -1 and giant components forming as thresholds approach the median vegetation fraction.

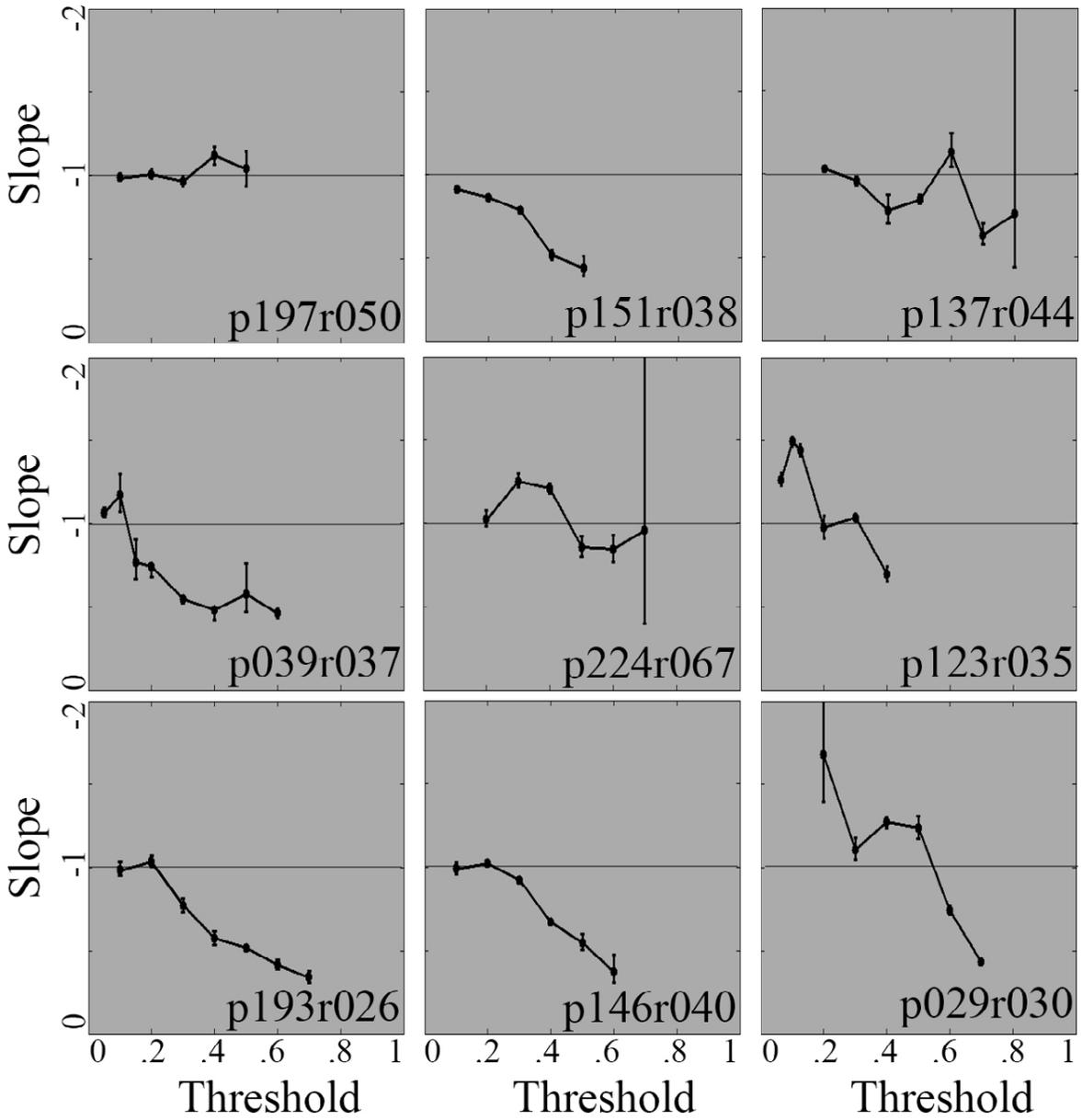

Figure 3c. Slope of Rank-Size distribution versus threshold for nine Landsat scenes. Local landscape properties vary from scene to scene, resulting in a wide range of vegetation fraction distributions. These distributions control the progression of slope of the size distributions.

### b. IKONOS

Figure 4 shows the procedure of successive thresholding when repeated for a 39 km$^2$ IKONOS image in Anhui, China. The 4-band image was unmixed into SVD fractions using local endmembers. Successive thresholding was then applied to the $F_v$ image. Segment area maps for four representative thresholds demonstrating the progression of the network are shown in the top 4 panels. The progression of the IKONOS size distributions with changing threshold (bottom right panel) is similar to that of the Landsat scene shown in Figure 2. At high thresholds IKONOS size distributions have high curvature and shallow slopes. The slope of the size distribution steepens as the threshold is reduced and the lower-tail power law cutoff gradually moves up the distribution. Curvature is even more pronounced than for Landsat at this phase. Once a threshold near 0.3 is reached, however, the size distribution loses most of its curvature and becomes linear. The slope of the size distribution crosses -1 at this point and the lower-tail cutoff rapidly moves deep into the lower tail of the distribution. As the threshold is decreased below this level, the network superconnects into a few giant components. The total number of segments (i.e. maximum rank) begins to decrease and the bottom of size distribution moves to the left. These properties are all similar to those observed for the 9 Landsat scenes in Figure 3.

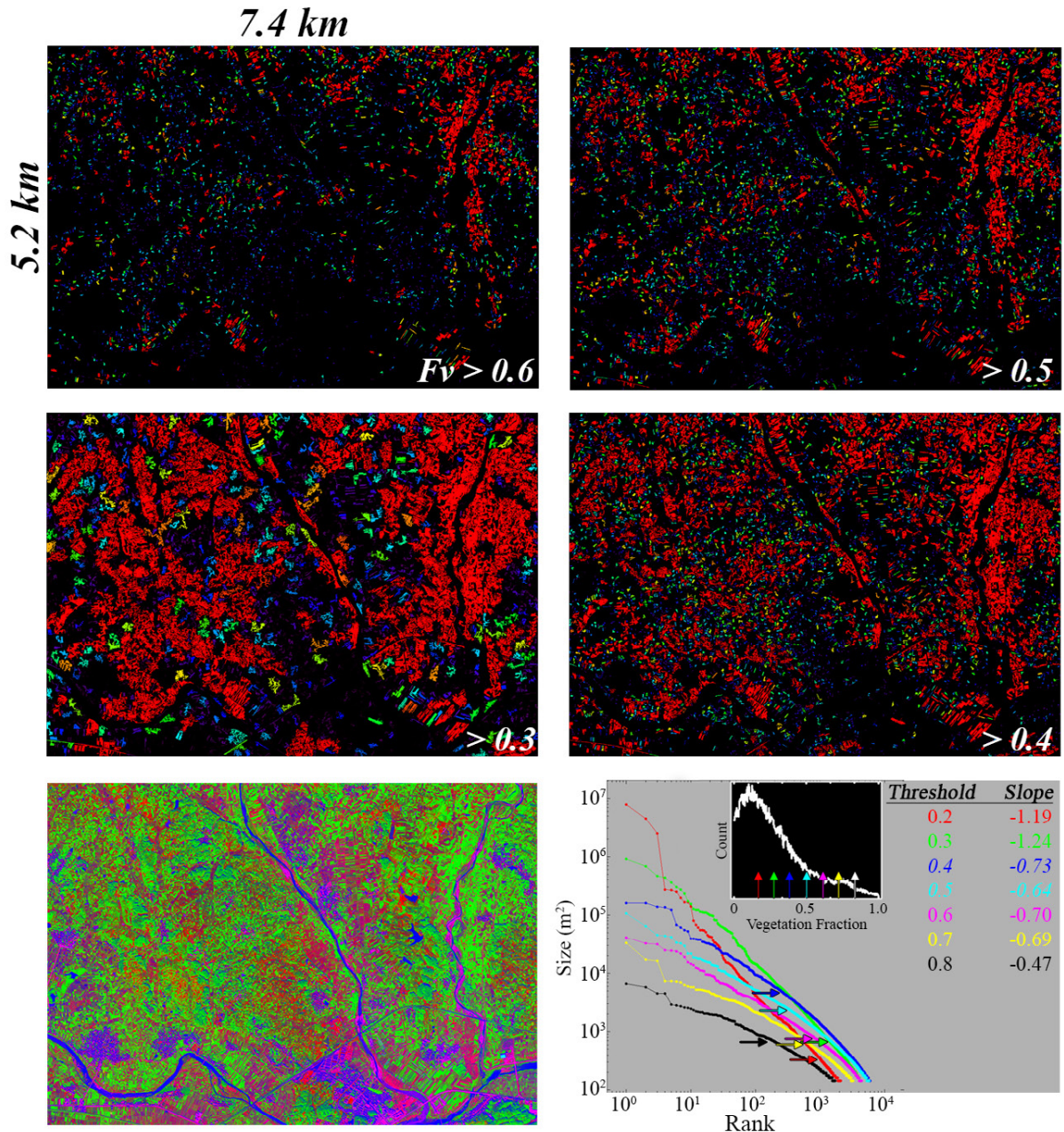

Figure 4. Successive thresholding of vegetation fraction for a 39 km² IKONOS image of Anhui, China. Rank-Size plots show a similar succession to those from Landsat in Figure 2.

### c. *Practical Example – Disruption by Node Removal*

Figure 5 shows how two agricultural networks can respond differently to disruption by sequentially reducing the area of each component. In each iteration of this process, all segments in the image are simultaneously reduced in size by removing one pixel width from around the boundary. We refer to this type of disruption as "erosion". We disrupt two agricultural networks in this way: one in the Salton Trough (p039r037) and one in South Dakota (p029r030). The upper tails of the rank-size distributions are shown in detail for successive numbers of erosional steps. The Salton Trough network (top) maintains the structure of its rank-size distribution through 7 erosional iterations, while the largest segments in the South Dakota network (bottom) rapidly dissociate into components with area approximately 2 orders of magnitude smaller, resulting in a drastic shallowing of the slope of the size distribution. This is a consequence of the differences in spatial structure and fractal dimension of the two networks.

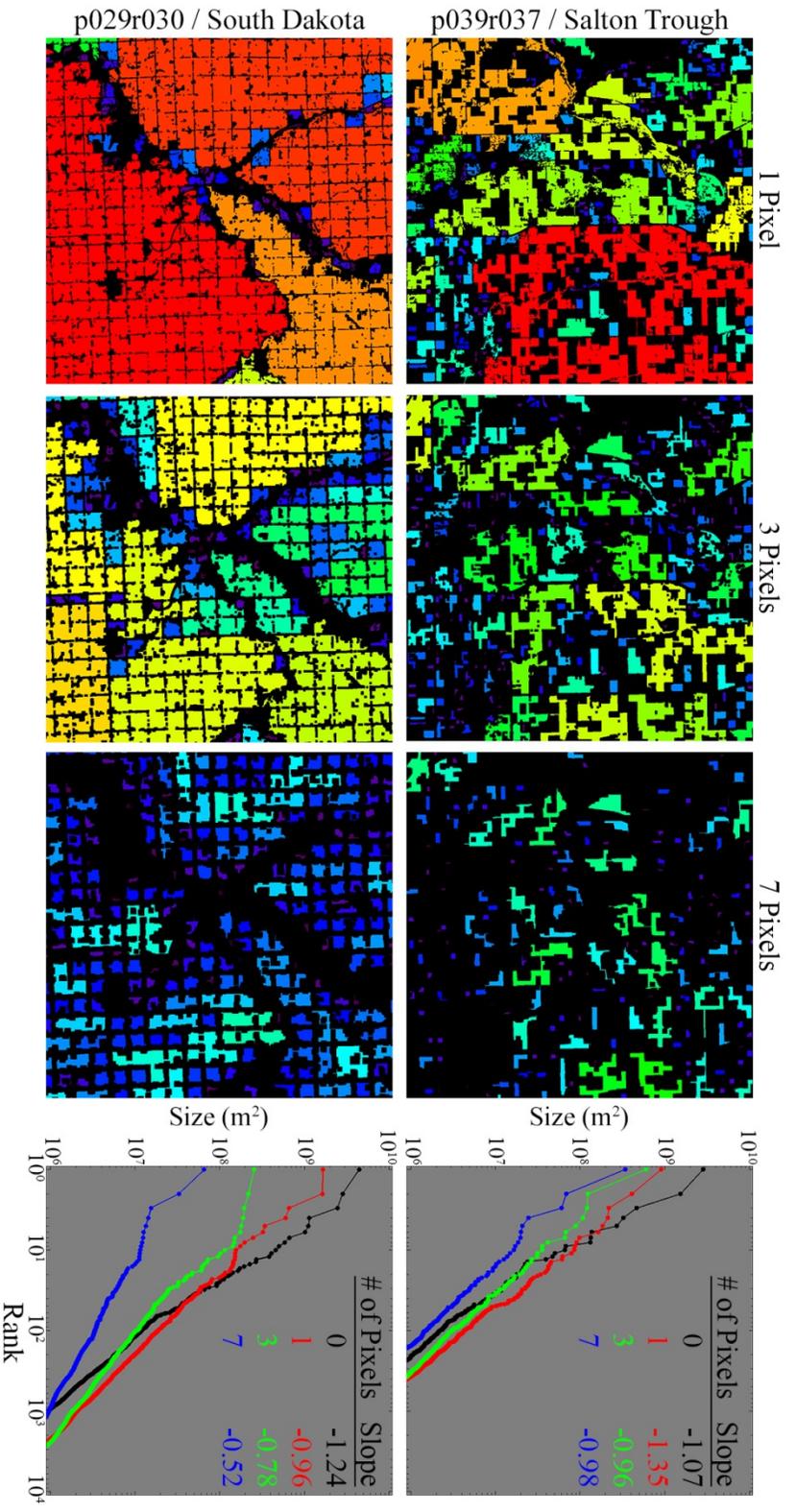

Figure 5. Disruption of agricultural networks by erosion. Images show a 1000 x 1000 pixel subset of segment area maps derived from full Landsat scenes. In each step, all boundary pixels are removed from the network, akin to removing rings of an onion. Regular rectilinear patterns correspond to real features of the landscape (i.e. roads) which often serve to guide the erosion process. As pixels are removed, the upper tail of the size distribution may maintain a slope near -1 (top) or flatten considerably (bottom), depending on the spatial structure of the network. Networks with size distributions which maintain their slope in the face of erosional perturbations may be more robust to disruption. Whether this form of network stability is desirable or undesirable depends on the application.

*Discussion*

Considerable range exists in the slope and curvature of the size distributions shown in Figure 3b – but the similarities are much more surprising than the differences. Indeed, we find it remarkable that there is any similarity at all given the diversity of landscapes (Figure 3a) and of vegetation abundance distributions (histogram insets of Figure 3b) from which they are derived. While it is clear that none of the 9 size distributions here exactly resembles the global size distribution in Figure 1, it is similarly clear that none of the 9 landscapes used in this study remotely resembles the diversity or scope of agriculture at global scales. Furthermore, because the differences between size distributions emerge from the differences in landscapes, these differences can be diagnostic in characterizing the variety in spatial distributions of agriculture across widely variable landscapes. From a network perspective, a diversity of size distributions implies a diversity of network structures.

Some of the size distributions in Figure 3b cannot plausibly be described as power laws. Some exhibit power law behavior that truncates in the middle of the distribution. Others show statistically plausible power law behavior extending deep into the lower tail of the distribution. We suggest that the important characteristic of the size distributions is not presence or absence of statistically defensible power law behavior, but rather that every distribution shown here is similarly heavy tailed. Every size distribution shows many more small patches than large patches, and nearly all distributions show that patches become both smaller and more frequent at similar rates, implying the total area sum of patches at any size is nearly equal to the total sum at any other size. This property corresponds to a slope of -1 on the plots in Figure 3b.

Further, Figure 3c shows that many of the distributions vary with threshold in a predictable way: starting at high threshold (right side of the plots), the size distribution increases in slope as the threshold drops and the components grow (moving right to left on the plot) until reaching linearity near -1. At this point, a giant component emerges and dominates the network. As the threshold is dropped even further, more and more of the remaining patches become connected into the giant component, reducing the total number of segments until every pixel in the entire domain is superconnected. The variations in progression of network structure with threshold are related to the fraction distributions, but the gross structure described above occurs in a consistent way across a wide range of conditions. A similar progression is also shown for the IKONOS image (Figure 4) over a much smaller spatial domain. Similar progressions have even been observed in random spatial networks and a general mechanism for the process been proposed (Small and Sousa 2015b). Despite this observed commonality, some of the distributions shown here vary with changing threshold in a more complex way than described above. This discrepancy often corresponds to severe non-Gaussianity in the $F_v$ histogram. Detailed analysis of this complexity will be the subject of further study.

Analysis of two seemingly similar agricultural landscapes by network erosion shown in Figure 5 demonstrates one potential application of the concepts presented in this paper. In one case (Salton Trough), power law behavior with slope near -1 is persistent even after removal of many pixels and considerable reduction of the total size of the network. In another case (South Dakota), the power law behavior of the network is much less robust. Removal of only a few pixels drastically reduces the sizes of the largest components (by a factor of ~100), rapidly breaking apart the largest segments of the network into much smaller disconnected components. This is clearly a result of the sectioning of the landscape by the regular grid of the road network.

One could imagine a landscape which is more sensitive to small perturbations as being more easily disruptable – either a dangerous characteristic (as in the case of pollinator pathways) or a desirable one (as in the case of containing disease outbreaks). Understanding the robustness of the structure of an agricultural network to disruption could provide application-specific insight into practical methods for disrupting (or preventing disruption of) connectivity across an agricultural landscape.

Another possible application, not shown in this analysis, is to use multitemporal observations to constrain the growth and attenuation of agricultural networks in a landscape throughout the complete phenological cycle. As the agricultural mosaic evolves through time, different crops are planted, green up, senesce, and are harvested at different times of year. Taken together, the combination of the spatial distribution of these crops and their corresponding phenology time series govern the complete spatiotemporal agricultural network of a landscape. The diagnostic property of an agricultural landscape may be not just the network as observed at any one time but rather the robustness of the network properties throughout the course of the year. For instance, effective pollination may require an agricultural network to remain in a particularly interconnected state for a certain length of time. Crops may be particularly susceptible to disease outbreaks at one particular time of year. Native species may be more sensitive to disruptions of habitat in migration season than at other times of year. Furthermore, network adaptation to catastrophic environmental stresses such as drought or widespread disease outbreaks may be easily characterized. Finally, multitemporal network studies – like all of the analyses performed in this paper – have the added benefit of being easily performed nearly anywhere on Earth using simple methodologies and freely available remotely sensed observations.

## Acknowledgements

D. Sousa is funded by a National Defense Science and Engineering Graduate Fellowship (NDSEG) through the U.S. Department of Defense. D. Sousa thanks F. Sousa. CS was funded by the NASA LCLUC program (grant LCLUC09-1-0023) and Interdisciplinary Science program (grants NNX12AM89G & NNN13D876T) and by the NASA Socioeconomic Data and Applications Center (SEDAC) (contract NNG13HQ04C). The authors thank Aaron Clauset and colleagues for providing the power law fitting code and methodology.